# Distributed Conceptual Structures

Robert E. Kent

The Ontology Consortium
rekent@ontologos.org

The theory of distributed conceptual structures, as outlined in this paper, is concerned with the distribution and conception of knowledge. It rests upon two related theories, Information Flow and Formal Concept Analysis, which it seeks to unify. Information Flow (IF) [2] is concerned with the distribution of knowledge. The foundations of Information Flow is explicitly based upon a mathematical theory known as the Chu Construction in *-autonomous categories [1] and implicitly based upon the mathematics of closed categories [5]. Formal Concept Analysis (FCA) [3] is concerned with the conception and analysis of knowledge [3]. In this paper we connect these two studies by extending the basic theorem of Formal Concept Analysis to the distributed realm of Information Flow. The main results are the categorical equivalence between classifications and concept lattices at the level of functions, and the categorical equivalence between bonds and complete adjoints at the level of relations. With this we hope to accomplish a rapprochement between Information Flow and Formal Concept Analysis.

## 1  Introduction

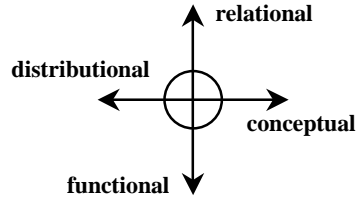

**Fig. 1. Two Dimensions of Distributed Conceptual Structures**

Figure 1 illustrates the conceptual structure of conceptual knowledge as a two-dimensional structure. The first dimension is along the distribution/conception distinction. Information Flow exists on the distributional side, whereas Formal Concept Analysis extends this toward the conceptual direction. The second dimension is along a functional/relational distinction. All of the development of Information Flow has taken place on the functional level, but some of Formal Concept Analysis extends this into the relational direction, which still might be considered *terra incognita*.

To a large extent the foundation of distributed conceptual structures is based upon binary relations (or matrices) and centered upon the axiom of adjointness between composition and residuation. This composition/residuation adjointness axiom is similar to the axiom of adjointness between conjunction and implication. Since composition and residuation are binary, the axiom has two statements:

- Left composition is (left) adjoint to left residuation:
  $r \circ s \subseteq t$ iff $s \subseteq r \backslash t$, for any compatible binary relations $r$, $s$ and $t$.
- Right composition is (left) adjoint to right residuation:
  $r \circ s \subseteq t$ iff $r \subseteq t/s$, for any compatible binary relations $r$, $s$ and $t$.

Some derived properties are that residuation preserves composition: $(r_1 \circ r_2) \backslash t = r_2 \backslash (r_1 \backslash t)$ and $t/(s_1 \circ s_2) = (t/s_2)/s_1$ and that residuation preserves identity: $Id_A \backslash t = t$ and $t/Id_A = t$. The involutions of transpose and negation are of secondary importance. The axiom for transpose states that transpose dualizes residuation: $(r \backslash t)^\smile = t^\smile / r^\smile$ and $(t/s)^\smile = s^\smile \backslash t^\smile$.

There are two important associative laws – one unconstrained the other constrained. These is an unconstrained associative law: $(r \backslash t)/s = r \backslash (t/s)$, for all $t \subseteq A \times B$, $r \subseteq A \times C$ and $s \subseteq D \times B$. There is also an associative law constrained by closure: if $t$ is an endo-relation and $r$ and $s$ are closed with respect to $t$, ($r = t/(r\backslash t)$ and $s = (t/s)\backslash t$) then $(t/s)\backslash r = s/(r\backslash t)$, for all $t \subseteq A \times A$, $r \subseteq A \times B$ and $s \subseteq C \times A$. Functions have a special behavior with respect to derivation. If function $f$ and relation $r$ are composable, then $f^{\infty}\backslash r = f \circ r$. If relation $s$ and the opposite of function $g$ are composable, then $s/g = s \circ g^{\infty}$.

The paper consists of five sections: Introduction, Basic Notions, Architecture, Limit/Colimit Constructions, and Summary and Future Work. The section on Basic Notions is a review of some of the basic ideas of Information Flow and Formal Concept Analysis (classifications, concept lattices, and functional infomorphisms), and an introduction of some new ideas (relational infomorphisms, bonds, and bonding pairs). The section on Architecture, the central section of the paper, describes the details of the conceptual structure of conceptual structures, and is principally concerned with the three categorical equivalences between the distributional pole and the conceptual pole. The section on Limit/Colimit Constructions, the closest to applications, gives an enhanced fibrational description of the architecture, thereby situating various basic constructions. The final section gives a summary and points out future plans to apply the logic of distributed conceptual structures to actual distributed representational frameworks on the Internet.

## 2 Basic Notions

There are six basic notions in distributed conceptual structures: classifications, concept lattices, functional infomorphisms, relational infomorphisms, bonds, and bonding pairs. The first two are object-like structures, with classifications being the central object-like notion of Information Flow, and concept lattices being the central object-like notion of Formal Concept Analysis. The last four are morphism-like, with functional infomorphisms being the central morphism notion of Information Flow, relational infomorphisms being newly defined in this paper, bonds being used as an analytic tool in Formal Concept Analysis, and bonding pairs being equivalent (in a categorical sense) to complete homomorphisms, the central morphism notion in FCA.

### 2.1 Objects

**Classifications.** According to the theory of Information Flow [2], information presupposes a system of classification. Classifications have been important in library science for the last 2,000 years. Major classification systems in library science include the Dewey Decimal System (DDS) and the Library of Congress (LC). The library science classification system most in accord with the philosophy and techniques of IF is the Colon classification system invented by the library scientist Ranganathan. A domain-neutral notion of classification is given by the following abstract mathematical definition. A classification $A = \langle inst(A), typ(A), \vDash_A \rangle$ consists of
1. a set, $inst(A)$, of things to be classified, called the *instances* of $A$,
2. a set, $typ(A)$, of things used to classify the instances, called the *types* of $A$, and
3. a binary *classification* relation, $\vDash_A$, between $inst(A)$ and $typ(A)$.

The notation $a \vDash_A \alpha$ is read "instance $a$ is of type $\alpha$ in $A$." Define the following pair of *derivation* operators: $A^A = \{\alpha \in typ(A) \mid a \vDash_A \alpha \text{ for all } a \in B\}$ for any instance subset $A \subseteq inst(A)$, and $\Gamma^A = \{b \in inst(A) \mid b \vDash_A \alpha \text{ for all } \alpha \in \Gamma\}$ for any type subset $\Gamma \subseteq typ(A)$. When $A : inst(A) \to 1$ and $\Gamma : typ(A) \leftarrow 1$ are regarded as relations, derivation is seen to be residuation, $A^A = A\backslash A$ and $\Gamma^A = A/\Gamma$, where the classification relation is simplified as $A$. An *extent* of $A$ is a subset of instances of the form $\Gamma^A$, and an *intent* of A is a subset of types of the form $A^A$. For any instance $a \in inst(A)$, the *intent* or *type set* of $a$ is the set $typ(a) = a^A = \{\alpha \in typ(A) \mid a \vDash_A \alpha\}$. Intent induces a preorder on the instances $inst(A)$ defined by: $a \leq_A a'$ when $a^A \supseteq a'^A$. For any type $\alpha \in typ(A)$, the *extent* or *instance set* of $\alpha$ is the set $inst(\alpha) = \alpha^A = \{a \in inst(A) \mid a \vDash_A \alpha\}$. Extent induces a preorder on types $typ(A)$ defined by: $\alpha \leq_A \alpha'$ when $\alpha^A \subseteq \alpha'^A$. Note that $a \subseteq \alpha^A$ iff $\alpha \subseteq a^A$, $\forall$ instance $a \in inst(A)$ and $\forall$ type $\alpha \in typ(A)$. A classification has an alternate expression in either a suitable extension of the theory of closed categories [5] or *-autonomous categories [1]: a *classification* is a triple $A = \langle inst(A), typ(A), \vDash_A \rangle$, where $inst(A)$ and $typ(A)$ are sets and $\vDash_A : typ(A) \times inst(A) \to 2$ is a function, representing a *2*-valued matrix. Classifications are known as *formal contexts* in Formal Concept Analysis [3]. In FCA, types are called *attributes* and instances are called *objects*.

As befitting such an important and generic notion, classifications abound. Organisms (instances) are classified by scientists in categories (types), such as Plant, Animal, Fungus, Bacterium, Alga, Eukaryote, Prokaryote, etc. Words (instances) are classified in a dictionary by parts of speech (types), such as Noun, Verb, Adjective, Adverb, etc. The following is a motivating example in Barwise and Seligman [2]: Given a first-order language *L*, the *truth classification* of *L* has *L*-structures as instances, sentences of *L* as types, and classification relation defined by $M \vDash \varphi$ if and only if sentence $\varphi$ is true in structure *M*. An ontology forms a classification with either explicit subtyping or subsumption being the classification relation. Any preorder $P = \langle P, \leq_P \rangle$ is a classification $P = \langle P, P, \leq_P \rangle$, where the preorder elements function as both types and instances, and the ordering relation is the classification.

Systemic examples of classifications also abound. Given any set *A* (of instances), the *instance powerset classification* $\wp A = \langle A, \wp A, \in \rangle$ associated with *A* has elements of *A* as instances and subsets of *A* as types with the membership relation serving as the classification relation. Given any classification $A = \langle inst(A), typ(A), \vDash_A \rangle$, the *dual classification* $A^\infty = \langle typ(A), inst(A), \vDash_A^\infty \rangle$ is the involution of *A*. This is the classification, whose instances are types of *A*, whose types are instances of *A*, and whose classification is the transpose of the *A* classification. The involution operator applies also to morphisms of classifications and limiting constructions on classifications.

**Concept Lattices.** The basic notion of FCA is the notion of a formal concept. A *formal concept* is a pair $\boldsymbol{a} = (A, \Gamma)$ of subsets, $A \subseteq inst(A)$, called the *extent* of the concept $\boldsymbol{a}$ and denoted $ext(\boldsymbol{a})$, and $\Gamma \subseteq typ(A)$, called the *intent* of the concept $\boldsymbol{a}$ and denoted $int(\boldsymbol{a})$, that satisfies the closure properties $A = \Gamma^A$ and $\Gamma = A^A$. There is a naturally defined concept order: $\boldsymbol{a}_1 \leq \boldsymbol{a}_2$ when $\boldsymbol{a}_1$ is more specific than $\boldsymbol{a}_2$ or dually when $\boldsymbol{a}_2$ is more generic than $\boldsymbol{a}_1$: $ext(\boldsymbol{a}_1) \subseteq ext(\boldsymbol{a}_2)$, or equivalently, $int(\boldsymbol{a}_1) \supseteq int(\boldsymbol{a}_2)$. This partial order is part of a complete lattice called the *concept lattice* of *A*, and denoted by $\boldsymbol{L}(A) = \langle L(A), \leq_A \rangle$, where the meet and join are defined by: $\boldsymbol{a}_1 \wedge_A \boldsymbol{a}_2 = (ext(\boldsymbol{a}_1) \cap ext(\boldsymbol{a}_2), (ext(\boldsymbol{a}_1) \cap ext(\boldsymbol{a}_2))^A)$ and $\boldsymbol{a}_1 \vee_A \boldsymbol{a}_2 = ((int(\boldsymbol{a}_1) \cap int(\boldsymbol{a}_2))^A, int(\boldsymbol{a}_1) \cap int(\boldsymbol{a}_2))$.

Define the *instance embedding relation* $\iota_A : inst(A) \to L(A)$, as follows: for every instance $a \in inst(A)$ and every formal concept $\boldsymbol{a} \in L(A)$ the relationship $a \iota_A \boldsymbol{a}$ holds when $a \in ext(\boldsymbol{a})$, *a* is in the extent of $\boldsymbol{a}$. This relation is closed on the right with respect to lattice order. Instances are mapped into the lattice by the *instance embedding function* $\iota_A : inst(A) \to L(A)$, where $\iota_A(a) = (\{a\}^{AA}, \{a\}^A)$ for each instance $a \in inst(A)$. This function is expressed in terms of the relation as the meet $\iota_A(a) = \wedge a \iota_A$. Concepts in $\iota_A[inst(A)]$ are called *instance concepts*. Any concept in the lattice $L(A)$ can be expressed as the join of a subset of instance concepts – $\iota_A[inst(A)]$ is join-dense in $L(A)$. Dually, define the *type embedding relation* $\tau_A : L(A) \to typ(A)$, as follows: for every type $\alpha \in typ(A)$ and every formal concept $\boldsymbol{a} \in L(A)$ the relationship $\boldsymbol{a} \tau_A \alpha$ holds when $\alpha \in int(\boldsymbol{a})$, $\alpha$ is in the intent of $\boldsymbol{a}$. This relation is closed on the left respect to lattice order. Types are mapped into the lattice by the *type embedding function* $\tau_A : typ(A) \to L(A)$, where $\tau_A(\alpha) = (\{\alpha\}^A, \{\alpha\}^{AA})$ for each type $\alpha \in typ(A)$. This function is expressed in terms of the relation as the join $\tau_A(\alpha) = \vee \tau_A \alpha$. Concepts in $\tau_A[typ(A)]$ are called *type concepts*. Any concept in the lattice $L(A)$ can be expressed as the meet of a subset of type concepts – $\tau_A[typ(A)]$ is meet-dense in $L(A)$. Any classification $a \vDash_A \alpha$ can be expressed in terms of the instance/type mappings as: $\iota_A(a) \leq_A \tau_A(\alpha)$ – the classification relation decomposes as the relational composition: $\vDash_A = \iota_A \circ \leq_A \circ \tau_A^{op}$.

The quintuple $\boldsymbol{L}(A) = \langle L(A), inst(A), typ(A), \iota_A, \tau_A \rangle$, is called the *concept lattice* of the classification *A*. More abstractly, a *concept lattice* $\boldsymbol{L} = \langle L, inst(L), typ(L), \iota_L, \tau_L \rangle$ consists of a complete lattice *L*, two sets $inst(L)$ and $typ(L)$ called the *instance set* and *type set* of $\boldsymbol{L}$, respectively; along with two functions mapping to the lattice, the *instance embedding* $\iota_L : inst(L) \to L$ and *type embedding* $\tau_L : typ(L) \to L$, such that $\iota_L[inst(L)]$ is join-dense in *L* and $\tau_L[typ(L)]$ is meet-dense in *L*. For each classification *A* the associated quintuple $\boldsymbol{L}(A)$ is a concept lattice.

Let *A* be any classification and let *A* be any (indexing) set. A *collective A-instance* indexed by *A* is any relation $a : inst(A) \to A$, and a *collective A-type* indexed by *A* is any relation $\alpha : A \to typ(A)$. Let $[inst(A), A]$ denote the preorder of *A*-indexed *A*-instances, and let $[A, typ(A)]$ denote the preorder of *A*- indexed $A$-types. The equivalence $\alpha \leq a\backslash A$ iff $a \circ \alpha \leq A$ iff $a \leq A/\alpha$ states that derivation with respect to $A$ forms an adjoint pair (Galois connection) $(\ )\backslash A \dashv A/(\ ) : [inst(A), A] \rightleftarrows [A, typ(A)]^{op}$. This observation can be extended to the statement that derivation forms a natural transformation $\S_A : \textbf{\textit{typ}}_A \Rightarrow \textbf{\textit{inst}}_A : \textbf{Relation}^{op} \to \textbf{Adjoint}$ between two functors between **Relation**$^{op}$ the opposite of the category of sets and binary relations and **Adjoint** the category of preorders and adjoint pairs of monotonic functions. Define the functor $\textbf{\textit{typ}}_A$ to map sets $A$ to the preorder of collective $A$-types $[A, typ(A)]^{\infty}$ and relations $r : A \leftarrow B$ to the adjoint pair $r\backslash(\ ) \dashv r\circ(\ )$, and define the functor $\textbf{\textit{inst}}_A$ to map sets $A$ to the preorder of collective $A$-instances $[inst(A), A]$ and relations $r : A \leftarrow B$ to the adjoint pair $(\ )\circ r \dashv (\ )/r$. For each set $X$ let $\S_{A, X}$ denote the derivation adjoint pair $(\ )\backslash A \dashv A/(\ )$. Then $\S_A$ is a natural transformation with $X$-th component $\S_{A, X}$.

An *A-indexed collective A-concept* is a pair $(a, \alpha)$, where $a : inst(A) \to A$ is a collective $A$-instance, $\alpha : A \to typ(A)$ is a collective $A$-type, which satisfy the closure conditions $a = A/\alpha$ and $\alpha = a\backslash A$. Let $A \triangleright A$ denote the collection (lattice) of all $A$-indexed collective $A$-concepts. The basic example of a collective A-concept is the pair $(\iota_A, \tau_A)$ consisting of the instance relation $\iota_A : inst(A) \to L(A)$ and the type relation $\tau_A : L(A) \to typ(A)$, since the closure conditions $\iota_A\backslash A = \tau_A$ and $A/\tau_A = \iota_A$ hold. This collective concept is indexed by the concept lattice $L(A)$. Any $A$-indexed collective $A$-concept $(a, \alpha)$ induces a unique *mediating function* $f : A \to L(A)$ satisfying the constraints $\alpha = f \circ \tau_A = f^{\infty}\backslash \tau_A$ and $a = \iota_A \circ f^{\infty} = \iota_A/f$. The definition $f(x) \triangleq (ax, x\alpha)$ is well-defined, since the closure conditions are equivalent to the (pointwise) fact that $(ax, x\alpha) \in L(A)$ is a formal concept for each indexing element $x \in A$. Conversely, for any function $f : A \to L(A)$, the pair $(\iota_A \circ f^{\infty}, f \circ \tau_A)$ is an $A$-indexed collective $A$-concept, since the conditions $\iota_A \circ f^{\infty} = A/(f \circ \tau_A) = (A/\tau_A)/f = \iota_A/f$ and $f \circ \tau_A = (\iota_A \circ f^{\infty})\backslash A = f^{\infty}\backslash(\iota_A\backslash A) = f^{\infty}\backslash \tau_A$ hold. So, we have the isomorphism $A \triangleright A \cong L(A)^A$, representing the fact that any $A$-indexed collective $A$-concept can equivalently be define as a function $f : A \to L(A)$.

The left adjoint $r\backslash(\ )$ preserves intents: if $(b, \beta)$ is a $B$-indexed collective $A$-concept, then $(A/((b\circ r)\backslash A), r\backslash \beta)$ is an $A$-indexed collective $A$-concept, since $r\backslash \beta = r\backslash(b\backslash A) = (b\circ r)\backslash A$, an $A$-indexed $A$-intent; in particular, if $\beta$ is a $B$-indexed collective $A$-intent, then $r\backslash \beta$ is an $A$-indexed collective $A$-intent. The right adjoint $(\ )/r$ preserves extents: if $(a, \alpha)$ is an $A$-indexed collective $A$-concept, then $(a/r, (A/(r\circ \alpha))\backslash A)$ is a $B$-indexed collective $A$-concept, since $a/r = (A/\alpha)/r = A/(r\circ \alpha)$, a $B$-indexed $A$-extent; in particular, if $a$ is an $A$-indexed collective $A$-extent, then $a/r$ is a $B$-indexed collective $A$-extent. These two observations are concentrated in an adjoint pair of monotonic functions, $\langle \varphi_r, \psi_r \rangle : A \triangleright B \rightleftarrows A \triangleright A$, where the left adjoint $\varphi_r : A \triangleright B \leftarrow A \triangleright A$ is defined by $\varphi_r((b, \beta)) = (A/((b\circ r)\backslash A), r\backslash \beta)$ and the right adjoint $\psi_r : A \triangleright B \to A \triangleright A$ is defined by $\psi_r((a, \alpha)) = (a/r, (A/(r\circ \alpha))\backslash A)$. Adjointness follows from the expressions $\alpha \leq r\backslash \beta$ iff $r\circ \alpha \leq \beta$ implies $b \leq a/r$ and the expressions $b \leq a/r$ iff $b\circ r \leq a$ implies $\alpha \leq r\backslash \beta$.

## 2.2 Morphisms

**Functional Infomorphisms.** Classifications are connected through infomorphisms. Infomorphisms connect classifications and provide a way to move information back and forth between classifications. There exist two generality levels of infomorphism: functional infomorphisms and relational infomorphisms. Functional infomorphisms are the infomorphisms that are used in the literature on Information Flow [2] and *-autonomous categories and the Chu construction [1]. Relational infomorphisms are newly defined in this paper. This subsection gives the definition of functional infomorphism. The next subsection will define the relational version.

A (*functional*) *infomorphism* $\langle f, g \rangle : A \rightleftarrows B$ from classification $A$ to classification $B$ is a contravariant pair of functions, a function $g : typ(A) \to typ(B)$ in the forward direction between types and a function $f : inst(A) \leftarrow inst(B)$ in the reverse direction between instances, satisfying the following fundamental property $f(b) \vDash_A \alpha$ iff $b \vDash_B g(\alpha)$ for each instance $b \in inst(B)$ and each type $\alpha \in typ(A)$. In the theory of the Chu construction [1] an infomorphism (a morphism in the category *Chu*(**Set**,*2*)) is known as a Chu transformation.

Systemic examples of functional infomorphisms abound. For any two sets (of instances) $A$ and $B$, any function $f : A \leftarrow B$ and its inverse image function $f^{-1} : \wp A \to \wp B$ form a *powerset infomorphism* $\wp f = \langle f, f^{-1} \rangle : \wp A \rightleftharpoons \wp B$ between the instance powerset classifications. Given any classification $A = \langle inst(A), typ(A), \vDash_A \rangle$, the *instance infomorphism* $\eta_A : A \rightleftharpoons \wp(inst(A))$, from $A$ to the powerset classification of the instance set of $A$, is instance-identity and defined on types as the instance set $\eta_A(\alpha) = inst(\alpha)$.

Given any two infomorphisms $f : A \rightleftharpoons B$ and $g : B \rightleftharpoons C$, there is a *composite infomorphism* $f \circ g : A \rightleftharpoons C$ defined by composing the type and instance functions. Given any classification $A = \langle inst(A), typ(A), \vDash_A \rangle$, the pair of identity functions on types and instances forms an *identity infomorphism* $Id_A : A \rightleftharpoons A$ (with respect to composition). Given any infomorphism $f : A \rightleftharpoons B$, the *dual infomorphism* is the infomorphism $f^\infty : B^\infty \rightleftharpoons A^\infty$, whose instance function is the type function of $f$ and whose type function is the instance function of $f$. Classifications and functional infomorphisms form a category **Classification** with involution $(\ )^\infty$.

The instance function is a monotonic function between instance preorders $f : inst(A) \leftarrow inst(B)$, and the type function is a monotonic function between type preorders $g : typ(A) \to typ(B)$. These facts are represented as two (projection) functors: There are two projection functors: the *instance functor* **inst** (contravariant) from **Classification** to **Preorder**[op], the opposite of the category of preorders and monontonic functions, maps each classification to its instance preorder and each infomorphism to its instance monotonic function; and the *type functor* **typ** from **Classification** to **Preorder** maps each classification to its type preorder and each infomorphism to its type monotonic function.

The fundamental property of functional infomorphisms is clearly related to the notion of adjointness. A pair of *adjoint monotonic functions* $\langle f, g \rangle : P \rightleftharpoons Q$ from preorder $P = \langle P, \leq_P \rangle$ to preorder $Q = \langle Q, \leq_Q \rangle$ is a contravariant pair of functions, a monotonic function $g : P \to Q$ in the forward direction and a monotonic function $f : P \leftarrow Q$ in the reverse direction, satisfying the following fundamental adjointness property $f(q) \leq_P p$ iff $q \leq_Q g(p)$ for each element $q \in P$ and each element $p \in P$. Preorders and adjoint pairs of monotonic functions form the category **Adjoint**. Projecting out to the left and right adjoint monotonic functions give rise to two (projection) functors, a *left* functor from **Adjoint** to **Preorder**[op], and a *right* functor from **Adjoint** to **Preorder**. Any adjoint pair of monotonic functions $\langle f, g \rangle : P \rightleftharpoons Q$ is a functional infomorphism between the associated classifications. This fact is expressed as an inclusion functor from **Adjoint** to **Classification** that commutes with projection functors.

**Relational Infomorphisms.** A (*relational*) *infomorphism* $\langle r, s \rangle : A \rightleftharpoons B$ from a classification $A = \langle inst(A), typ(A), \vDash_A \rangle$ to classification $B = \langle inst(B), typ(B), \vDash_B \rangle$ is a contravariant pair of binary relations, a *type relation* $s : typ(A) \to typ(B)$ between types and an *instance relation* $r : inst(A) \to inst(B)$ between instances, satisfying the following fundamental property: $r \backslash A = B / s$. This property is equivalent to either of the following properties:

$$\alpha \in (rb)^A \text{ iff } b \in (\alpha s)^B \qquad \forall_{b \in inst(B)} \text{ and } \forall_{\alpha \in typ(A)}$$
$$rb \subseteq \alpha^A \text{ iff } \alpha s \subseteq b^B$$
$$rB \subseteq \Gamma^A \text{ iff } \forall_{b \in B, \alpha \in \Gamma} \, rb \subseteq \alpha^A \text{ iff } \forall_{b \in B, \alpha \in \Gamma} \, \alpha s \subseteq b^B \text{ iff } \Gamma s \subseteq B^B \qquad \forall_{B \subseteq inst(B)} \text{ and } \forall_{\Gamma \subseteq typ(A)}$$
$$r \circ B \subseteq A/\Gamma \text{ iff } B \circ \Gamma \subseteq r \backslash A \text{ iff } B \circ \Gamma \subseteq B/s \text{ iff } \Gamma \circ s \subseteq B \backslash B \qquad \forall_{B : inst(B) \to 1} \text{ and } \forall_{\Gamma : 1 \to typ(A)}$$

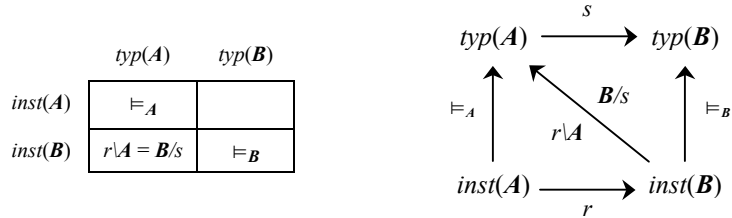

|  | $typ(A)$ | $typ(B)$ |
|---|---|---|
| $inst(A)$ | $\vDash_A$ |  |
| $inst(B)$ | $r \backslash A = B/s$ | $\vDash_B$ |

The common relation $[r \backslash A = B/s] : inst(B) \to typ(A)$ in the fundamental property is called the *bond* of $\langle r, s \rangle$.

Given any two relational infomorphisms $\langle r_1, s_1 \rangle : A \rightleftharpoons B$ and $\langle r_2, s_2 \rangle : B \rightleftharpoons C$, there is a *composite infomorphism* $\langle r_1, s_1 \rangle \circ \langle r_2, s_2 \rangle = \langle r_2 \circ r_1, s_1 \circ s_2 \rangle : A \rightleftharpoons C$ defined by composing the type and instance relations, and whose fundamental property follows from the preceding composition and associative laws. Given any classification $A = \langle inst(A), typ(A), \vDash_A\rangle$, the pair of identity functions on types and instances, with the bond being the classification relation, forms an *identity infomorphism* $Id_A : A \rightleftharpoons A$ (with respect to composition). For any given infomorphism $\langle r, s\rangle : A \rightleftharpoons B$ the *dual infomorphism* $\langle r, s\rangle^\infty = \langle s^\infty, r^\infty\rangle : B^\infty \rightleftharpoons A^\infty$ is the infomorphism with type and instance relations switched and transposed. The fundamental property of relational infomorphisms for composition, identity and involution follow from basic properties of residuation. Classifications and relational infomorphisms form a category **Classification**$_{rel}$ with involution $(\ )^\infty$.

Any functional infomorphism $\langle f, g\rangle : A \rightleftharpoons B$ has an associated relational infomorphism **fn2rel**$(\langle f, g\rangle) = \langle f_\bullet, g^\bullet\rangle : A \rightleftharpoons B$, whose bond is the relation in the fundamental property $[f(b) \vDash_A \alpha$ iff $b \vDash_B g(\alpha)] : inst(B) \to typ(A)$. The definition of the type and instance relations uses the induced orders on instances and types: the type relation $g^\bullet : typ(A) \to typ(B)$ is defined by $g^\bullet(\alpha, \beta) = g(\alpha) \leq_B \beta$, and the instance relation $f_\bullet : inst(A) \to inst(B)$ is defined by $f_\bullet(a, b) = a \leq_A f(b)$. The operator **fn2rel**, which maps classifications to themselves, is a functor from **Classification** to **Classification**$_{rel}$.

The theory of classifications with relational infomorphisms can profitably be regarded as a theory of Boolean matrices, with classifications being matrices, infomorphisms matrix pairs, composition involving matrix multiplication in the two dual senses of relational composition and residuation, and involution using matrix transpose.

**Bonds.** A *bond* [3] $F : A \to B$ between two classifications $A$ and $B$ is a classification $F = \langle inst(B), typ(A), \vDash_F\rangle$, sharing types with $A$ and instances with $B$, that is compatible with $A$ and $B$ in the sense of closure: type sets $\{bF \mid b \in inst(B)\}$ are intents of $A$ and instance sets $\{F\alpha \mid \alpha \in typ(A)\}$ are extents of $B$. Closure can be expressed categorically as $(A/F)\backslash A = F$ and $B/(F\backslash B) = F$. The first expression says that $(A/F, F)$ is an $inst(B)$-indexed collective $A$-concept, whereas the second says that $(F, F\backslash B)$ is a $typ(A)$-indexed collective $B$-concept. A bond is order-closed on left and right: $b' \leq_B b$, $bF\alpha$ imply $b'F\alpha$, and $bF\alpha$, $\alpha \leq_A \alpha'$ imply $bF\alpha'$. Or, $b' \leq_B b$ implies $b'F \supseteq bF$, and $\alpha \leq_A \alpha'$ implies $F\alpha \subseteq F\alpha'$.

For any two bonds $F : A \to B$ and $G : B \to C$, the composition is the bond $F \circ G \triangleq (B/G)\backslash F : A \to C$ defined using left and right residuation. Since both $F$ and $G$ being bonds are closed with respect to $B$, an equivalent expression for the composition is $F \circ G \triangleq G/(F\backslash B) : A \to C$. Pointwise, the composition is $F \circ G = \{(c, \alpha) \mid F\alpha \supseteq (cG)^B\}$. To check closure, $(A/(F \circ G))\backslash A = (A/((B/G)\backslash F))\backslash A = (B/G)\backslash F$, since $F$ being a collective $A$-intent means that $(B/G)\backslash F$ is also a collective $A$-intent. With respect to bond composition, the identity bond at any classification $A$ is the classification relation $\vDash_A : A \to A$; its closed relational infomorphism is the instance-type order pair $\langle \leq_A, \leq_A\rangle : A \to A$. With bond composition and bond identities, classifications and bonds form the category **Bond**.

For any relational infomorphism $\langle r, s\rangle : A \rightleftharpoons B$ the common relation **bond**$(\langle r, s\rangle) = r\backslash A = B/s$ in the fundamental property is a bond: for any instance $b \in inst(B)$ the $b$-th row of the bond $r\backslash A$ is an intent of $A$, since $b(r\backslash A) = \{\alpha \in typ(A) \mid rb \subseteq \alpha^A\} = (rb)^A$, and for any type $\alpha \in typ(A)$, the $\alpha$-th column of the bond $B/s$ is an extent of $B$, since $(B/s)\alpha = \{b \in inst(B) \mid \alpha s \subseteq b^B\} = (\alpha s)^B$. Any bond $F : A \to B$ is the bond of some (closed) relational infomorphism $\langle r, s\rangle : A \rightleftharpoons B$ – just make the definitions: $r \triangleq A/F$ and $s \triangleq F\backslash B$, or pointwise $rb \triangleq (bF)^A$ and $\alpha s \triangleq (F\alpha)^B$. There is a naturally defined equivalence relation on the relational infomorphisms between any two classifications $A$ and $B$: two infomorphisms are equivalent $\langle r, s\rangle \equiv \langle r', s'\rangle$ when they have the same bond. Since the bond of a composition is the composition of the bonds and the bond of the identity is the identity bond, this defines a **bond** quotient functor from relational infomorphisms **Classification**$_{rel}$ to bonds **Bond**, that makes the category **Bond** a quotient category of **Classification**$_{rel}$.

The notion of a bond of a relational infomorphism is (ignoring orientation) the same notion of a bond as defined in Ganter and Wille [3]. More to the point, the notion of a relational infomorphism, defined for the first time in this paper, support a categorical rendering of the notion of bond as defined in Ganter and Wille [3].

**Bonding Pairs.** A *complete* (*lattice*) *homomorphism* $\psi : L \to K$ between complete lattices $L$ and $K$ is a (monotonic) function that preserves both joins and meets. Being meet-preserving, $\psi$ has a left adjoint $\varphi : K \to L$, and being join-preserving $\psi$ has a right adjoint $\theta : K \to L$. So, a complete homomorphism is the middle monotonic function in two adjunctions $\varphi \dashv \psi \dashv \theta$. Let **Complete Lattice** denote the category of complete lattices and complete homomorphisms.

The bond equivalent to a complete homomorphism would seem to be given by two bonds $F : A \to B$ and $G : B \to A$ where the right adjoint $\psi_F : L(A) \to L(B)$ of the complete adjoint $A(F) = \langle \varphi_F, \psi_F \rangle : L(A) \rightleftharpoons L(B)$ of one bond (say $F$, without loss of generality) is equal to the left adjoint $\varphi_G : L(A) \to L(B)$ of the complete adjoint $A(G) = \langle \varphi_G, \psi_G \rangle : L(B) \rightleftharpoons L(A)$ of the other bond $G$ with the resultant adjunctions, $\varphi_F \dashv \psi_F = \varphi_G \dashv \psi_G$, where the middle adjoint is the complete homomorphism. This is indeed the case, but the question is what constraint to place on $F$ and $G$ in order for this to hold. The simple answer is to identify the actions of the two monotonic functions $\psi_G$ and $\varphi_F$, and this is exactly the solution given in Ganter and Wille [3]. Let $(A, \Gamma) \in L(B)$ be any formal concept in $L(A)$. The action of the left adjoint $\varphi_G$ on this concept is $(A, \Gamma) \mapsto (A^{GB}, A^G)$, whereas the action of the right adjoint $\psi_F$ on this concept is $(A, \Gamma) \mapsto (\Gamma^F, \Gamma^{FB})$. So the appropriate pointwise constraints are: $A^{GB} = \Gamma^F$ and $\Gamma^{FB} = A^G$, for every concept $(A, \Gamma) \in L(A)$. We now give these pointwise constraints a categorical rendition.

A *bonding pair* $\langle F, G \rangle : A \rightleftharpoons B$ between two classifications $A$ and $B$ is a contravariant pair of bonds, a bond $F : A \to B$ in the forward direction and a bond $G : B \to A$ in the reverse direction, satisfying the following *pairing constraints*:

$$F/\tau_A = B/(\iota_A \backslash G) \text{ and } \iota_A \backslash G = (F/\tau_A) \backslash B, \qquad (1)$$

which state that $(F/\tau_A, \iota_A \backslash G)$ is an $L(A)$-indexed collective $B$-concept. The definitions of the relations $F/\tau_A$ and $\iota_A \backslash G$ are given as follows: $F/\tau_A = \{(b, a) \mid int(a) \subseteq bF\} = \{(b, a) \mid ((bF)^A, bF) \leq_B a\}$ and $\iota_A \backslash G = \{(a, \beta) \mid ext(a) \subseteq G\beta\} = \{(a, \beta) \mid a \leq_B (G\beta, (G\beta)^A)\}$. Any concept $a = (A, \Gamma) \in L(A)$ is mapped by the relations as: $(F/\tau_A)((A, \Gamma)) = \{b \mid \Gamma \subseteq bF\} = \Gamma^F$ and $(\iota_A \backslash G)((A, \Gamma)) = \{\beta \mid A \subseteq G\beta\} = A^G$. Hence, pointwise the constraints are $\Gamma^F = B^{GB}$ and $A^G = \Gamma^{FB}$. These are the original pointwise constraints of Ganter and Wille [3] discussed above.

The pointwise constraints can be lifted to a collective setting – any bonding pair $\langle F, G \rangle : A \rightleftharpoons B$ preserves collective concepts: for any $A$-indexed collective $A$-concept $(a, \alpha)$, $a \backslash A = \alpha$ and $A/\alpha = a$, the conceptual image $(F/\alpha, a \backslash G)$ is an $A$-indexed collective $B$-concept, $B/(a \backslash G) = F/\alpha$ and $(F/\alpha) \backslash B = a \backslash G$. An important special case is the $L(A)$-indexed collective $A$-concept $(\iota_A, \tau_A)$. To state that the $\langle F, G \rangle$-image $(F/\tau_A, \iota_A \backslash F)$ is an $L(A)$-indexed collective $B$-concept, is to assert the pairing constraints $F/\tau_A = B/(\iota_A \backslash G)$ and $\iota_A \backslash G = (F/\tau_B) \backslash B$. So, the concise definition in terms of pairing constraints, the original pointwise definition of Ganter and Wille [3], and that assertion that $\langle F, G \rangle$ preserves all collective concepts, are equivalent versions of the notion of a bonding pair.

Two other special cases are the collective $B$-concepts $(F, A \backslash G)$ and $(F/A, G)$ with pairing constraints $B/(A \backslash G) = F$ and $(F/A) \backslash B = G$, which are $\langle F, G \rangle$-images of the collective $A$-concepts $(\vDash_A, \leq_A)$ and $(\leq_A, \vDash_A)$. Since any collective concept uniquely factors in terms of its mediating function, the image of any collective concept can be computed in two steps: (1) Factor an $A$-indexed collective $A$-concept $(a, \alpha)$ in terms of the mediating function $f : A \to L(A)$ and the basic collective A-concept $(\iota_A, \tau_A)$; and (2) Compose the $\langle F, G \rangle$-image $(F/\tau_A, \iota_B \backslash G)$ with the mediating function $f$, resulting in the $\langle F, G \rangle$-image $A$-indexed collective $A$-concept $(F/\alpha, a \backslash G)$.

Let $\langle F, G \rangle : A \rightleftharpoons B$ and $\langle M, N \rangle : B \rightleftharpoons C$ be two bonding pairs. Define the bonding pair composition $\langle F, G \rangle \circ \langle M, N \rangle \triangleq \langle F \circ M, N \circ G \rangle : A \rightleftharpoons C$ in terms of bond composition. We can check, either categorically or pointwise, that bonding pair composition is well defined. Let **Bonding Pair** denote the category, whose objects are classifications and whose morphisms are bonding pairs.

## 3  Architecture

The architecture of the distribution/conception distinction is a categorical equivalence at both the functional and the relational poles of the other (scope) distinction. This architecture is in one sense a categorical expression of the basic theorem of FCA. The central architecture is revealed at the functional level to be the equivalence between **Classification** and **Concept Lattice**, at the relational level to be the equivalence between **Bond** and **Complete Adjoint**, and at the complete relational level to be the equivalence between **Bonding Pair** and **Complete Lattice.** Not to be forgotten is the fact from Information Flow, that **Classification** is also equivalent to **Regular Theory**, the category of regular theories and theory morphisms.

### 3.1 Functional Equivalence

Information Flow (IF) and Formal Concept Analysis (FCA) are intimately connected. Every classification supports, and is equivalent to, an associated complete lattice called its concept lattice. Every infomorphism defines an adjoint pair of monotonic functions between the concept lattices of its source and target classifications. This section formalizes these observations in a theorem on categorical equivalence.

**The Concept Lattice Functor.** Let $\langle f, g \rangle : A \rightleftharpoons B$ be any functional infomorphism between two classifications with instance function $f : inst(B) \rightarrow inst(A)$ and type function $g : typ(A) \rightarrow typ(B)$. How are the two concept lattices $L(A)$ and $L(B)$ related?

Since for any concept $(A, \Gamma) \in L(A)$ the equality $f^{-1}[A] = g[\Gamma]^B$ holds between direct and inverse images, the mapping $(A, \Gamma) \mapsto (f^{-1}[A], (f^{-1}[A])^B)$ is a well-defined monotonic function $L(f) : L(A) \rightarrow L(B)$ that preserves type concepts in the sense that: $\tau_A \cdot L(f) = g \cdot \tau_B$. Since it is always true that meet-irreducible concepts are type concepts, if $L(B)$ is type reduced, then $L(f)$ preserves meet-irreducibility – it maps meet-irreducible concepts to meet-irreducible concepts. Dually, since for any concept $(B, \Delta) \in L(B)$ the equality $f[B]^A = g^{-1}[\Delta]$ holds between direct and inverse images, the mapping $(g^{-1}[\Delta]^A, g^{-1}[\Delta]) \mapsto (B, \Delta)$ is a well-defined monotonic function $L(g) : L(A) \leftarrow L(B)$ that preserves instance concepts in the sense that $\iota_B \cdot L(g) = f \cdot \iota_A$. Since it is always true that join-irreducible concepts are instance concepts, if $L(A)$ is instance reduced, then $L(g)$ preserves join-irreducibility – it maps join-irreducible concepts to join-irreducible concepts. Moreover, $\langle L(g), L(f) \rangle : L(A) \rightleftharpoons L(B)$ is a pair of adjoint monotonic functions between concept lattices. The quadruple $L(\langle f, g \rangle) = \langle L(g), L(f), f, g \rangle$ is called the *concept lattice morphism* of the infomorphism $\langle f, g \rangle : A \rightleftharpoons B$.

More abstractly, a *concept lattice morphism* $\langle \varphi, \psi, f, g \rangle : L \rightleftharpoons K$ between two concept lattices $L = \langle L, inst(L), typ(L), \iota_L, \tau_L \rangle$ and $K = \langle K, inst(K), typ(K), \iota_K, \tau_K \rangle$ consists of a pair of ordinary functions $f : inst(L) \leftarrow inst(K)$ and $g : typ(L) \rightarrow typ(K)$ between instance sets and type sets, respectively; and a pair $\langle \varphi, \psi \rangle : L \rightleftharpoons K$ of adjoint monotonic functions, where the right adjoint $\psi : L \rightarrow K$ is a monotonic function in the forward direction that preserves types $\tau_L \cdot \psi = g \cdot \tau_K$, and the left adjoint $\varphi : L \leftarrow K$ is a monotonic function in the reverse direction that preserves instances $\iota_K \cdot \varphi = f \cdot \iota_L$. Let **Concept Lattice** denote the category of concept lattices and concept lattice morphisms.

For each infomorphism $\langle f, g \rangle : A \rightleftharpoons B$ the quadruple $L(\langle f, g \rangle) : L(A) \rightleftharpoons L(B)$ is a concept lattice morphism from $L(A) = \langle L(A), inst(A), typ(A), \iota_A, \tau_A \rangle$ the concept lattice of classification $A$ to $L(B) = \langle L(B), inst(B), typ(B), \iota_B, \tau_B \rangle$ the concept lattice of classification $B$. The operator $L$ is a functor **Classification** $\rightarrow$ **Concept Lattice** called the *concept lattice functor* from the category of classifications and functional infomorphisms to the category of concept lattices and concept lattice morphisms.

**The Classification Functor.** Associated with any concept lattice $L = \langle L, inst(L), typ(L), \iota_L, \tau_L \rangle$ is the classification $C(L) = \langle inst(L), typ(L), \vDash_L \rangle$, which has $L$-instances as its instance set, $L$-types as its type set, and the relational composition $\vDash_L = \iota_L \circ \leq_L \circ \tau_L^{op}$ as its classification relation. Associated with any concept lattice morphism $\langle \varphi, \psi, f, g \rangle : L \rightleftarrows K$, from concept lattice $L$ to concept lattice $K$, is the infomorphism $C(\langle \varphi, \psi, f, g \rangle) = \langle f, g \rangle : C(L) \rightleftarrows C(K)$. The fundamental property of infomorphisms is an easy translation of the adjointness condition for $\langle \varphi, \psi \rangle : L \rightleftarrows K$ and the commutativity of the instance/type functions with the lattice monotonic functions. The operator $C$ is a functor **Concept Lattice** $\rightarrow$ **Classification**, from the category of concept lattices and their morphisms to the category of classifications and infomorphisms, called the *classification functor*.

**Equivalence.** The functor composition $C \circ L$ is naturally isomorphic to the identity functor $Id_{\text{Concept Lattice}}$. To see this, let $L = \langle L, inst(L), typ(L), \iota_L, \tau_L \rangle$ be a concept lattice with associated classification $C(L) = \langle inst(L), typ(L), \vDash_L \rangle$. Part of the fundamental theorem of concept lattices [3] asserts the isomorphism $L \cong L(C(L))$. In particular, define the map $L(C(L)) \rightarrow L$ by $(A, \Gamma) \mapsto \vee_L \iota_L(A) = \wedge_L \tau_L(\Gamma)$ for every formal concept $(A, \Gamma)$ in $L(C(L))$, and the map $L \rightarrow L(C(L))$ by $x \mapsto (\{a \in inst(L) \mid \iota_L(a) \leq_L x\}, \{\alpha \in typ(L) \mid x \leq_L \tau_L(\alpha)\})$ for every element $x \in L$. These are inverse monotonic functions. Let $\langle \varphi, \psi, f, g \rangle : L \rightleftarrows K$ be a concept lattice morphism between concept lattices $L = \langle L, inst(L), typ(L), \iota_L, \tau_L \rangle$ and $K = \langle K, inst(K), typ(K), \iota_K, \tau_K \rangle$ with associated infomorphism $C(\langle \varphi, \psi, f, g \rangle) = \langle f, g \rangle : C(L) \rightleftarrows C(K)$. Then, up to isomorphism, $L(C(\langle \varphi, \psi, f, g \rangle)) = \langle \varphi, \psi, f, g \rangle$. This defines the natural isomorphism: $C \circ L \cong Id_{\text{Concept Lattice}}$.

The functor composition $L \circ C$ is equal to the identity functor $Id_{\text{Classification}}$. To see this, consider whether $A = C(L(A))$ for any classification $A$. Obviously, the type and instance sets are the same. What about the classification relations? The classification relation in $C(L(A))$ is defined in terms of the lattice order and instance/type embeddings by $\vDash_L = \iota_L \circ \leq_L \circ \tau_L^{op}$; which is easily see to be equal $\vDash_A$. Hence, $A = C(L(A))$. What about infomorphisms? The functional infomorphisms $\langle f, g \rangle$ and $C(L(\langle f, g \rangle))$ are equal.

### 3.2 Relational Equivalence

Relational infomorphisms are more general than functional infomorphisms. This increased flexibility and expressiveness must be balanced with a decreased number of properties. However, the property of (categorical) equivalence between the distributional side and the conceptual side still holds. This section formalizes these observations in a theorem on categorical equivalence.

**The Complete Adjoint Functor.** Let $F : A \rightarrow B$ be any bond and let $\langle r, s \rangle : A \rightleftarrows B$ be any relational infomorphism having $F$ as its bond. We again ask how the two concept lattices $L(A)$ and $L(B)$ related, but now in terms of relational infomorphisms. More particularly, how are $L(A)$ and $L(B)$ related to $L(F)$, the concept lattice of the bond itself?

Since for any concept $(B, \Gamma) \in L(F)$ the intent $\Gamma = B^F = \cap_{b \in B} bF$ is an intent of $L(A)$, the mapping $(B, \Gamma) \mapsto (\Gamma^A, \Gamma)$ is a well-defined monotonic function $\partial_0 : L(F) \rightarrow L(A)$ that has a right-adjoint-right-inverse $\tilde{\partial}_0 : L(A) \rightarrow L(F)$ defined as the mapping $(A, \Gamma) \mapsto (\Gamma^F, \Gamma^{FF})$. Let $int_F = \langle \partial_0, \tilde{\partial}_0 \rangle : L(A) \rightleftarrows L(F)$ denote this adjoint pair. Dually, since for any concept $(B, \Gamma) \in L(F)$ the extent $B = \Gamma^F = \cap_{\alpha \in \Gamma} F\alpha$ is an extent of $L(B)$, the mapping $(B, \Gamma) \mapsto (B, B^B)$ is a well-defined monotonic function $\partial_1 : L(F) \rightarrow L(B)$ that has a left-adjoint-right-inverse $\tilde{\partial}_1 : L(B) \rightarrow L(F)$ defined as the mapping $(B, \Delta) \mapsto (B^{FF}, B^F)$. Let $ext_F = \langle \tilde{\partial}_1, \partial_1 \rangle : L(F) \rightleftarrows L(A)$ denote this adjoint pair.

Since for any concept $(A, \Gamma) \in L(A)$ the equality $\Gamma^F = F/\Gamma = (B/s)/\Gamma = B/(\Gamma \circ s) = (\Gamma s)^B$ holds, the mapping $(A, \Gamma) \mapsto (\Gamma^F, \Gamma^{FB})$ is a well-defined monotonic function $L(s) : L(A) \rightarrow L(B)$, which is the composition $L(s) = \tilde{\partial}_0 \cdot \partial_1$. Dually, since for any concept $(B, \Delta) \in L(B)$ the equality $B^F = B\backslash F = B\backslash(r\backslash A) = (r \circ B)\backslash A = (rB)^A$ holds, the mapping $(B, \Delta) \mapsto (B^{FA}, B^F)$ is a well-defined monotonic function $L(r) : L(A) \leftarrow L(B)$, which is the composition $L(r) = \tilde{\partial}_1 \cdot \partial_0$. Basic properties of residuation show that this is an adjoint pair of monotonic functions $\langle L(r), L(s) \rangle = int_F \circ ext_F : L(A) \rightleftarrows L(B)$, since $B^F \supseteq \Gamma$ iff $B\backslash F \supseteq \Gamma$ iff $B \circ \Gamma \subseteq F$ iff $B \subseteq F/\Gamma$ iff $B \subseteq \Gamma^F$. The pair $A(F) \triangleq \langle L(r), L(s) \rangle$ is called the *complete adjoint* of the bond $F : A \to B$.

More abstractly, a *complete adjoint* $\langle \varphi, \psi \rangle : L \rightleftarrows K$ between two complete lattices $L = \langle L, \leq_L, \wedge_L, \vee_L \rangle$, and $K = \langle K, \leq_K, \wedge_K, \vee_K \rangle$ consists of a pair $\langle \varphi, \psi \rangle : L \rightleftarrows K$ of adjoint monotonic functions. Let **Complete Adjoint** denote the category of complete lattices and adjoint monotonic functions. For each bond $F : A \to B$ the pair $A(F) : L(A) \rightleftarrows L(B)$ is a complete adjoint from $L(A)$ the concept lattice of classification $A$ to $L(B)$ the concept lattice of classification $B$. The operator $A$ is a functor **Bond** $\to$ **Complete Adjoint** called the *complete adjoint functor* from the category of classifications and bonds to the category of complete lattices and adjoint pairs.

**The Bond Functor.** Associated with any complete lattice $L = \langle L, \leq_L, \wedge_L, \vee_L \rangle$ is the classification $B(L) = \langle L, L, \leq_L \rangle$, which has $L$-elements as its instances and types, and the lattice order as its classification relation. Associated with any adjoint pair $\langle \varphi, \psi \rangle : L \rightleftarrows K$, from complete lattice $L$ to complete lattice $K$, is the bond $B(\langle \varphi, \psi \rangle) : B(L) \rightleftarrows B(K)$ defined by its adjointness property: $yB(\langle \varphi, \psi \rangle)x$ iff $\varphi(y) \leq_L x$ iff $y \leq_K \psi(x)$ for all elements $x \in L$ and $y \in K$. The closure property of bonds is obvious, since $yB(\langle \varphi, \psi \rangle) = \uparrow_L \varphi(y)$ for all elements $y \in K$ and $B(\langle \varphi, \psi \rangle)x = \downarrow_K \psi(x)$ for all elements $x \in L$. The operator $B$ is a functor **Complete Adjoint** $\to$ **Bond**, from the category of complete lattices and adjoint pairs to the category of classifications and bonds, called the *bond functor*

**Equivalence.** The functor composition $B \circ A$ is naturally isomorphic to the identity functor $Id_{\textbf{Complete Adjoint}}$. To see this, let $L = \langle L, \leq_L, \wedge_L, \vee_L \rangle$ be a complete lattice with associated classification $B(L) = \langle L, L, \leq_L \rangle$. Part of the fundamental theorem of concept lattices [3] asserts the isomorphism $L \cong A(B(L))$. In particular, formal concepts of $A(B(L))$ are of the form $(\downarrow_L x, \uparrow_L x)$ for elements $x \in L$. So, define the obvious maps $A(B(L)) \to L$ by $(\downarrow_L x, \uparrow_L x) \mapsto x$ and $L \to L(A(L))$ by $x \mapsto (\downarrow_L x, \uparrow_L x)$, for every element $x \in L$. Let $\langle \varphi, \psi \rangle : L \rightleftarrows K$ be a complete adjoint, an adjoint pair of monotonic functions, between complete lattices $L = \langle L, \leq_L, \wedge_L, \vee_L \rangle$, and $K = \langle K, \leq_K, \wedge_K, \vee_K \rangle$ with associated bond $B(\langle \varphi, \psi \rangle) : B(L) \to B(K)$. Then, the right adjoint of $A(B(\langle \varphi, \psi \rangle))$ maps $(\downarrow_L x, \uparrow_L x) \mapsto (\downarrow_K \psi(x), \uparrow_K \psi(x))$ and the left adjoint of $A(B(\langle \varphi, \psi \rangle))$ maps $(\downarrow_K y, \uparrow_K y) \mapsto (\downarrow_L \varphi(y), \uparrow_L \varphi(y))$. So, up to isomorphism, $A(B(\langle \varphi, \psi \rangle))$ is the same as $\langle \varphi, \psi \rangle$. This defines the natural isomorphism: $B \circ A \cong Id_{\textbf{Complete Adjoint}}$.

The basic theorem of Formal Concept Analysis [3] can be framed in terms of two fundamental bonds (relational infomorphisms) between any classification and its associated concept lattice. For any classification $A$ the instance embedding relation is a bond $\iota_A : L(A) \to A$ from the concept lattice to $A$ itself. The pair $\langle L(A)/\iota_A, \tau_A \rangle : L(A) \rightleftarrows A$ is a relational infomorphism whose bond is the instance embedding relation. For any classification $A$ the type embedding relation is a bond $\tau_A : A \to L(A)$ from $A$ to its concept lattice. The pair $\langle \iota_A, \tau_A | L(A) \rangle : A \rightleftarrows L(A)$ is a relational infomorphism whose bond is the type embedding relation. The instance and type embedding bonds are inverse to each other: $\iota_A \circ \tau_A = Id_{L(A)}$ and $\tau_A \circ \iota_A = Id_A$.

The functor composition $A \circ B$ is naturally isomorphic to the identity functor $Id_{\textbf{Bond}}$. To see this, let $A$ be a classification with associated concept lattice $A(A)$. The comments above demonstrate the isomorphism $A \cong B(A(A))$. Let $F : A \to B$ be a bond between classifications $A$ and $B$ with associated complete adjoint $A(F) : A(A) \rightleftarrows A(B)$. The bond $B(A(F)) : B(A(A)) \to B(A(B))$ contains a conceptual pair $(a, b)$ of the form $a = (A, \Gamma)$ and $b = (B, \Delta)$ iff $B \circ \Gamma \subseteq F$, where $B \circ \Gamma = B \times \Gamma$ a Cartesian product or rectangle, iff $B \subseteq \Gamma^F$ iff $\Gamma \subseteq B^F$. So, $(\iota_A \circ F) \circ \tau_B = (F/(\iota_A\backslash A)) \circ \tau_B = (F/\tau_A) \circ \tau_B = (B/\tau_B)\backslash(F/\tau_A) = \iota_B\backslash(F/\tau_A) = B(A(F))$, by bond composition and properties of the instance and type relations. Hence, $B(A(F)) \circ \iota_B = \iota_A \circ F$. This proves the required naturality condition.

### 3.3  Complete Relational Equivalence

In the section on relational equivalence, we have seen how bonds are categorically equivalent to complete adjoints, adjoint pairs between complete lattices. Unfortunately, these are not the best morphisms for making structural comparisons between complete lattices. Complete homomorphisms are best for this [3]. Since complete homomorphisms are special cases of complete adjoints on the conceptual side, we are interested in what constraints to place on bonds on the distributional side.

**The Complete Lattice Functor.** Let $\langle F, G \rangle : A \rightleftarrows B$ be any bonding pair. Then $F : A \to B$ is a bond in the forward direction from classification $A$ to classification $B$, and $G : A \leftarrow B$ is a bond in the reverse direction to classification $A$ from classification $B$. Applying the complete adjoint functor **A : Bond → Complete Adjoint**, we get two adjoint pairs in opposite directions: an adjoint pair $\langle \varphi_F, \psi_F \rangle : L(A) \rightleftarrows L(B)$ in the forward direction and an adjoint pair $\langle \varphi_G, \psi_G \rangle : L(B) \rightleftarrows L(A)$ in the reverse direction. It was shown above that for bonding pairs the meet-preserving monotonic function $\psi_F : L(A) \to L(B)$ is equal to the join-preserving monotonic function $\varphi_G : L(A) \to L(B)$, giving a complete homomorphism. This function is the unique mediating function for the $L(A)$-indexed collective $B$-concept $(F/\tau_A, \iota_A\backslash G)$, the $\langle F, G \rangle$-image of the $L(A)$-indexed collective $A$-concept $(\iota_A, \tau_A)$, whose closure expressions define the pairing constraints.

The *complete lattice functor* $\mathbf{A}^2$ **: Bonding Pair → Complete Lattice** is the operator that maps a classification $A$ to its concept lattice $\mathbf{A}^2(A) \triangleq L(A)$ regarded as a complete lattice only, and maps a bonding pair $\langle F, G \rangle : A \rightleftarrows B$ to its complete homomorphism $\mathbf{A}^2(\langle F, G \rangle) \triangleq \psi_F = \varphi_G : L(A) \to L(B)$.

**The Bonding Pair Functor.** Let $\psi : L \to K$ be a complete homomorphism between complete lattices $L$ and $K$ with associated adjunctions $\varphi \dashv \psi \dashv \theta$. Then $\langle \varphi, \psi \rangle : L \rightleftarrows K$ and $\langle \psi, \phi \rangle : K \rightleftarrows L$ are morphisms in **Complete Adjoint**. Application of the bond functor **B** produces the bonds $F = \mathbf{B}(\langle \varphi, \psi \rangle) : \mathbf{B}(L) \to \mathbf{B}(K)$ and $G = \mathbf{B}(\langle \psi, \theta \rangle) : \mathbf{B}(K) \to \mathbf{B}(L)$ between the classifications $\mathbf{B}(L) = \langle L, L, \leq_L \rangle$ and $\mathbf{B}(K) = \langle K, K, \leq_K \rangle$. Note that for any complete lattice $L$ the instance and type relations for the classification $\mathbf{B}(L)$ are both equal to the order relation $\iota_L = \leq_L = \tau_L$. Since $F/\tau_L = F/\leq_L = F$, $\mathbf{B}(K)/(\iota_L\backslash G) = \leq_K/(\leq_L\backslash G) = \leq_K/G = F$, $\iota_L\backslash G = \leq_L\backslash G = G$ and $(F/\tau_L)\backslash \mathbf{B}(K) = (F/\leq_L)\backslash\leq_K = F\backslash\leq_K = G$, the pair of bonds $(F, G)$ is a bonding pair $(F, G) = (\mathbf{B}(\langle \varphi, \psi \rangle), \mathbf{B}(\langle \psi, \phi \rangle)) : \mathbf{B}(L) \rightleftarrows \mathbf{B}(K)$. Let $\mathbf{B}^2(\psi)$ denote this pair.

The *bonding pair functor* $\mathbf{B}^2$ **: Complete Lattice → Bonding Pair** is the operator that maps a complete lattice $L$ to its classification $\langle L, L, \leq_L \rangle$ and maps a complete homomorphism to its bonding pair as above. Since the bond functor **B** is functorial, so is $\mathbf{B}^2$.

**Equivalence.** The functor composition $\mathbf{A}^2 \circ \mathbf{B}^2$ is naturally isomorphic to the identity functor $Id_{\text{Bonding Pair}}$. Consider any classification $A$. The type and instance embedding relations form bonding pairs in two different ways, $\langle \tau_A, \iota_A \rangle : A \rightleftarrows L(A)$ and $\langle \iota_A, \tau_A \rangle : L(A) \rightleftarrows A$, and these are inverse to each other: $\langle \tau_A, \iota_A \rangle \circ \langle \iota_A, \tau_A \rangle = Id_A$ and $\langle \iota_A, \tau_A \rangle \circ \langle \tau_A, \iota_A \rangle = Id_{L(A)}$. So that each classification is isomorphic in **Bonding Pair** to its concept lattice: $A \cong \mathbf{B}^2(L(A)) = \mathbf{A}^2 \circ \mathbf{B}^2(A)$. Let $\langle F, G \rangle : A \rightleftarrows B$ be a bonding pair. As shown above, the naturality conditions for bonds $F$ and $G$ are expressed as $\iota_A \circ F \circ \tau_B = \mathbf{A} \circ \mathbf{B}(F)$ and $\iota_B \circ G \circ \tau_A = \mathbf{A} \circ \mathbf{B}(G)$. So $\langle \iota_A, \tau_A \rangle \circ \langle F, G \rangle \circ \langle \tau_A, \iota_A \rangle = \langle \iota_A \circ F \circ \tau_B, \iota_B \circ G \circ \tau_A \rangle = \langle \mathbf{A} \circ \mathbf{B}(F), \mathbf{A} \circ \mathbf{B}(G) \rangle = \mathbf{A}^2 \circ \mathbf{B}^2(\langle F, G \rangle)$.

The functor composition $\mathbf{B}^2 \circ \mathbf{A}^2$ is naturally isomorphic to the identity functor $Id_{\text{Complete Lattice}}$. Consider any complete lattice $L$. As we have seen in studying relational equivalence, any complete lattice $L$ is isomorphic to the complete lattice $\mathbf{B}^2 \circ \mathbf{A}^2(L) = L(\langle L, L, \leq_L \rangle)$ via the bijection $x \leftrightarrow (\downarrow_L x, \uparrow_L x)$. Now consider any complete homomorphism $\psi : L \to K$ be a complete homomorphism between complete lattices $L$ and $K$ with associated adjunctions $\varphi \dashv \psi \dashv \theta$. The bonding pair functor maps this to the bonding pair $\mathbf{B}^2(\psi) = (\mathbf{B}(\langle \varphi, \psi \rangle), \mathbf{B}(\langle \psi, \phi \rangle))$, and the complete lattice functor maps this to the complete homomorphism $\mathbf{A}^2(\mathbf{B}^2(\psi)) = \tilde{\psi} : L(\langle L, L, \leq_L \rangle) \to L(\langle K, K, \leq_K \rangle)$ with associated adjunctions $\tilde{\varphi} \dashv \tilde{\psi} \dashv \tilde{\theta}$, where $\tilde{\varphi}((\downarrow_K y, \uparrow_K y)) = (\downarrow_L \varphi(y), \uparrow_L \varphi(y))$, $\tilde{\psi}((\downarrow_L x, \uparrow_L x)) = (\downarrow_K \psi(x), \uparrow_K \psi(x))$ and $\tilde{\theta}((\downarrow_K y, \uparrow_K y)) = (\downarrow_L \theta(y), \uparrow_L \theta(y))$. Clearly, the naturality condition holds between $\psi$ and $\mathbf{B}^2 \circ \mathbf{A}^2(\psi)$.

### 3.4 Theorem and Architectural Diagram

**Theorem.** *The distributional/conceptual distinction in the architecture of distributed conceptual structures is represented by three categorical equivalences.*

- *The category of classifications and functional infomorphisms is categorically equivalent to the category of concept lattices and concept lattice morphisms*

  **Classification ≡ Concept Lattice**

  *via the lattice functor and classification functor, which are generalized inverses:* $\mathbf{L} \circ \mathbf{C} = Id_{\text{Classification}}$ *and* $\mathbf{C} \circ \mathbf{L} \cong Id_{\text{Concept Lattice}}$.

- *The category of classifications and bonds is categorically equivalent to the category of complete lattices and complete adjoints*

  **Bond ≡ Complete Adjoint**

  *via the complete adjoint functor and bond functor, which are generalized inverses:* $\mathbf{A} \circ \mathbf{B} \cong Id_{\text{Bond}}$ *and* $\mathbf{B} \circ \mathbf{A} \cong Id_{\text{Complete Adjoint}}$.

- *The category of classifications and bonding pairs is categorically equivalent to the category of complete lattices and complete homomorphisms*

  **Bonding Pair ≡ Complete Lattice**

  *via the complete lattice functor and bonding pair functor, which are generalized inverses:* $\mathbf{A}^2 \circ \mathbf{B}^2 \cong Id_{\text{Bonding Pair}}$ *and* $\mathbf{B}^2 \circ \mathbf{A}^2 \cong Id_{\text{Complete Lattice}}$.

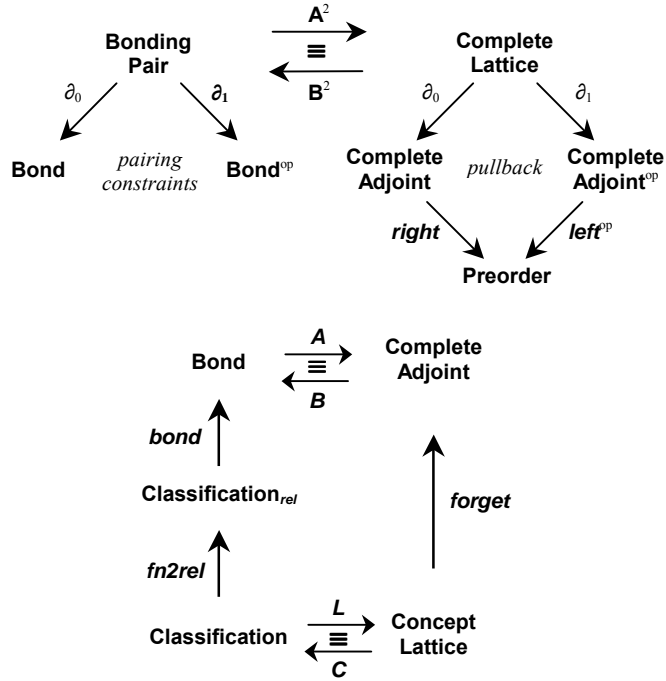

**Fig. 2. Architectural Diagram of Distributed Conceptual Structures**

Figure 2 contains a commuting diagram of functors, categories and equivalences that represents the architecture of distributed conceptual structures. This is the central contribution of this paper. Many of the details that support this diagram are known from the literature [2, 3]. This paper seeks to bring these facts together into a coherent view. Figure 2 is two-dimensional, having the same orientation as Figure 1. The vertical dimension contains the functional-relational distinction. The morphisms of the equivalent categories **Classification** and **Concept Lattice** are function-based, whereas the equivalent categories **Bond** and **Complete Adjoint** are relation-based. The horizontal dimension contains the distributional-conceptual distinction. This is represented as categorical equivalences. The equivalences between the categories **Classification** and **Concept Lattice** at the functional level and the categories **Bond** and **Complete Adjoint** at the relational level are expressions of the basic theorem of Formal Concept Analysis [3]. The equivalence between the categories **Classification** and **Regular Theory** is one of the main contributions of the theory of Information Flow [2].

## 4  Limit/Colimit Constructions

Limit/colimit constructions are very important in applications, often harboring the central semantics. Categorical equivalence is more general and more useful that categorical isomorphism. Equivalent categories have the same limit/colimit structures [see the appendix], and since the architecture diagram in Figure two has three categorical equivalence, limit/colimit structures will give added meaning to distributive conceptual structures. In brief comments below, although we describe limit/colimit constructions on one side of the equivalence, we are assured of their preservation when mapped to the other side of the equivalence.

A true conception of the limit/colimit architecture requires an understanding the two-dimensional fibrational nature of distributive conceptual structures. The fibrational (instance) dimension is defined in terms of the instance forgetful functor **inst** : **Classification** → **Set**$^{op}$, here simplified by ignoring instance order. For any set (of instances) $A$ the $A$-th fiber category of **inst**, denoted **inst**$^{-1}(A)$, is the subcategory of classifications $A$ with instance set $inst(A) = A$ and infomorphisms $\langle f, g \rangle : A \rightleftharpoons B$ with instance function $f = Id_A$. The fibrational type dimension is defined in terms of the type forgetful functor **typ** : **Classification** → **Set**. For any set (of types) $\Phi$ the $\Phi$-th fiber category of **typ**, denoted **typ**$^{-1}(\Phi)$, is the subcategory of classifications $B$ with type set $typ(B) = \Phi$ and infomorphisms with type function $Id_\Phi$. The initial object in the $A$-th fiber category **inst**$^{-1}(A)$ is $0_A = \langle A, \emptyset, \emptyset \rangle$, the terminal object in **inst**$^{-1}(A)$ is $\wp A = \langle A, \wp A, \in \rangle$ the *instance powerset classification over A*. For any classification $A$ the instance infomorphism $\eta_A : A \rightleftharpoons \wp(inst(A))$ is the unique infomorphism in **inst**$^{-1}(A)$ from $A$ to $\wp A$. Given two classifications $A_0$ and $A_1$ in the $A$-th fiber category **inst**$^{-1}(A)$, the apposition $A_0 | A_1$ is the coproduct of $A_0$ and $A_1$ in **inst**$^{-1}(A)$. Dual comments hold for the type fibrational dimension, where for example subposition is the product construction.

The full category **Classification** contains all limits and colimits. Given two classifications $A$ and $B$, the *sum* (*semiproduct*) $A+B$ is the coproduct of $A$ and $B$ in **Classification**. The dual notion provides the product. All limits/colimits exist when not only products/coproducts but also quotient constructions exist. Given a classification $A$, a *dual invariant* is a pair $J = (A, R)$ consisting of a set $A \subseteq inst(A)$ of instances of $A$ and a binary relation $R$ on types of $A$ satisfying the constraint: if $\alpha R \beta$, then for each $a \in A$, $a \vDash_A \alpha$ if and only if $a \vDash_A \beta$. The *dual quotient* of $A$ by $J$, written $A/J$, is the classification with instances $A$, whose types are the $R$-equivalence classes of types of $A$, and whose classification is $a \vDash_{A/J} [\alpha]$ if and only if $a \vDash_A \alpha$. Dual quotients include kernel image factorization, coequalizers and pushouts. Suitable colimit constructions in **Classification** have been used to define the semantics of ontology sharing [4].

## 5  Summary and Future Work

This paper has had as its goal the formulation of a conceptual framework for conceptual knowledge representation. For this it uses the language of category theory in order to represent some of the essence of Information Flow and Formal Concept Analysis, thereby unifying these two studies. This has culminated in the recognition of the two-dimensional nature of distributed conceptual structures (Figure 1), whose particulars are described as a fundamental commuting diagram of categories, functors and equivalence (Figure 2). Information Flow has developed the distributed nature of the logic of information, principally represented by the morphisms on the distributional side of the diagram. Formal Concept Analysis has developed the conceptual nature of knowledge, principally represented by the objects on the conceptual side of the diagram.

The author has been engaged for the last five years in the design of markup languages based upon the model of distributed conceptual structures at the functional level in Figure 2. The [Ontology Markup Language (OML)](#) had as its goal the representation of ontologically structured information corresponding roughly to **Classification**. Many ontologies have been designed using OML. [The Conceptual Knowledge Markup Language (CKML)](#) had as its goal the representation of conceptually structured information corresponding roughly to **Concept Lattice**. Also included in CKML were capabilities for conceptual scaling corresponding very roughly to the categorical equivalence between classifications and concept lattices. Current and future work is the design of a new markup language [Information Flow Framework (IFF)](#), a [distillation](#) of the essence of OML and CKML.

# 6 Appendix

According to Saunders Mac Lane [6] equivalences between categories are more general, and more useful, than isomorphisms between categories. We emphatically concur and we argue that the main reason for their usefulness comes from the fact that equivalent categories have the same limit/colimit structures. We make this precise in the following theorem.

**Fact.** *If categories* **A** *and* **B** *are equivalent, then they have the same limit/colimit structures. More particularly, if* **B** *has limits (colimits) for all* **C**-*shaped diagrams, then* **A** *also has limits (colimits) for all* **C**-*shaped diagrams, and vice-versa. In particular,* **A** *is complete (co-complete) iff* **B** *is complete (co-complete).*

**Proof.** Assume $F : \mathbf{A} \to \mathbf{B}$ and $G : \mathbf{B} \to \mathbf{A}$ are functors that mediate the equivalence through natural isomorphisms $\eta : Id_\mathbf{A} \Rightarrow F \cdot G$ and $\varepsilon : G \cdot F \Rightarrow Id_\mathbf{B}$. This means the following.

- $\eta$ is the unit and $\varepsilon$ is the counit of adjunction $F \dashv G$:   $\eta F \bullet F\varepsilon = Id_F$ and $G\eta \bullet \varepsilon G = Id_G$.
- The natural transformations $\eta$ are $\eta^{-1}$ are inverse:   $\eta \bullet \eta^{-1} = Id_{Id_\mathbf{A}}$ and $\eta^{-1} \bullet \eta = Id_{F \cdot G}$.
- The natural transformations $\varepsilon$ are $\varepsilon^{-1}$ are inverse:   $\varepsilon \bullet \varepsilon^{-1} = Id_{G \cdot F}$ and $\varepsilon^{-1} \bullet \varepsilon = Id_{Id_\mathbf{B}}$.
- $\varepsilon^{-1}$ the unit and $\eta^{-1}$ is the counit of adjunction $G \dashv F$:   $\varepsilon^{-1}G \bullet G\eta^{-1} = Id_G$ and $F\varepsilon^{-1} \bullet \eta^{-1}F = Id_F$.

Let $D : \mathbf{C} \to \mathbf{A}$ be any **C**-shaped diagram in **A**. Diagram $D$ is mapped by $F$ to $D \cdot F : \mathbf{C} \to \mathbf{B}$, a **C**-shaped diagram in **B**. By assumption, there is a limiting cone $\lambda : B \Rightarrow D \cdot F$ for $D \cdot F$ in **B** with limit object $B$. We will show that $\lambda G \bullet D\eta^{-1} : G(B) \Rightarrow D$ is a limiting cone for $D$ in **A** with limit object $G(B)$.

Let $\gamma : A \Rightarrow D$ be any cone for $D$ in **A**. Then $\gamma F : F(A) \Rightarrow D \cdot F$ is a cone for $D \cdot F$ in **B**. Since $\lambda$ is a limiting cone, there is a unique **B**-arrow $g : F(A) \to B$ with $g \cdot \lambda = \gamma F$. Define **A**-arrow $f \triangleq \eta_A \cdot G(g) : A \to G(F(A)) \to G(B)$. We will show that $f$ is the unique mediating **A**-arrow for cone $\lambda G \bullet D\eta^{-1}$; that is, $f$ is the unique **A**-arrow satisfying the constraint $f \cdot \lambda G \bullet D\eta^{-1} = \gamma$.

[Existence] $f \cdot G(\lambda_i) \cdot \eta^{-1}{}_{Di} = \eta_A \cdot G(g) \cdot G(\lambda_i) \cdot \eta^{-1}{}_{Di} = \eta_A \cdot G(g \cdot \lambda_i) \cdot \eta^{-1}{}_{Di} = \eta_A \cdot G(F(\gamma_i)) \cdot \eta^{-1}{}_{Di} = \eta_A \cdot \eta_A^{-1} \cdot \gamma_i = \gamma_i$. [Uniqueness] Suppose that $\tilde{f} : A \to G(B)$ is any **A**-arrow satisfying $\tilde{f} \cdot (\lambda G \bullet D\eta^{-1}) = \gamma$. Applying $F$, get $F(\tilde{f}) \cdot \eta^{-1}{}_B \cdot \lambda_i = F(\tilde{f}) \cdot F(G(\lambda_i)) \cdot \eta^{-1}{}_{F(Di)} = F(\tilde{f}) \cdot F(G(\lambda_i)) \cdot F(\eta^{-1}{}_{Di}) = F(\tilde{f} \cdot G(\lambda_i) \cdot \eta^{-1}{}_{Di}) = F(\gamma_i) = g \cdot \lambda_i$. By uniqueness $F(\tilde{f}) \cdot \eta_B^{-1} = F(\tilde{f}) \cdot \eta^{-1}{}_B = g$. Applying $G$, get $\tilde{f} = \eta_A \cdot G(F(\tilde{f})) \cdot G(\eta_B^{-1}) = \eta_A \cdot G(g)$.